\documentclass[a4paper]{jpconf}
\usepackage{graphicx}
 \def\BA{\begin{eqnarray}}
 \def\BE{\begin{equation}}
 \def\BF{\begin{figure}[htb]}
 \def\BT{\begin{table}[htb]}
 \def\EA{\end{eqnarray}}
 \def\EE{\end{equation}}
 \def\EF{\end{figure}}
 \def\ET{\end{table}}

 \def\la{\langle}
 \def\ra{\rangle}

 \def\lsim{\mathrel{\rlap{\lower4pt\hbox{\hskip1pt$\sim$}}
     \raise1pt\hbox{$<$}}}         
 \def\gsim{\mathrel{\rlap{\lower4pt\hbox{\hskip1pt$\sim$}}
     \raise1pt\hbox{$>$}}}         

\newcommand{\AmS}{{\protect\the\textfont2
  A\kern-.1667em\lower.5ex\hbox{M}\kern-.125emS}}

\begin{document}
\vspace*{-0.7cm}
\title{\hspace*{-0.2cm}
Nuclear suppression of dileptons at forward rapidities
}

\author{J. \v Cepila$^{1}$ and J. Nemchik$^{1,2}$}

\address{$^1$ 
Czech Technical University in Prague, FNSPE,
B\v rehov\' a 7, 11519 Prague, Czech Republic
}

\address{$^2$
Institute of Experimental Physics SAS,
Watsonova 47, 04001 Ko\v sice, Slovakia
}

\ead{jan.cepila@fjfi.cvut.cz;~~~
nemchik@saske.sk}

\begin{abstract}
Data from E772 and E866 experiments on the Drell-Yan process
exhibit a significant nuclear
suppression at large Feynman $x_F$.
We show that a corresponding kinematic
region does not allow to interpret this as a
manifestation of coherence or a Color Glass Condensate.
We demonstrate, however, that this suppression 
can be treated alternatively as an effective energy loss
proportional to initial energy.
To eliminate suppression coming from the coherence,
we perform predictions for nuclear effects also at large dilepton masses.
Our calculations are in a good agreement with available data.
Since the 
kinematic limit
can be also 
approached in 
transverse momenta $p_T$, we present
in the RHIC energy range
corresponding predictions for expected large-$p_T$ suppression as well.
Since a new experiment E906 planned at FNAL will provide us
with more precise data soon, we present also predictions for
expected large-$x_F$ nuclear suppression in this kinematic region.
\end{abstract}
\vspace*{-0.9cm}

\section{Introduction}

In comparison with a central region
of very small rapidities, $y\to 0$,
the forward rapidity region allows to study
processes corresponding to 
much higher initial energies accessible 
at mid rapidities.
If a particle with mass $M$ and transverse momentum
$p_T$ is produced in a hard reaction then the corresponding
values of Bjorken variable in the beam
and the target are
$x_{1,2} = \sqrt{M^2 + p_T^2}\,e^{\pm y}/\sqrt{s}$.
Thus, at forward rapidities the target $x_2$ is $e^y$-
times smaller than at mid rapidities. This allows
to study already at RHIC coherence phenomena (shadowing,
Color Glass Condensate (CGC)), which are expected
to suppress particle yields.

Forward rapidity physics,
manifested itself as a strong nuclear suppression,
has been already investigated in
variety of processes at different energies:
in production of different   
species of particles in $p+A$ collisions \cite{small-pt},
in charge pion
\cite{na49} and charmonium production
\cite{na3-jp,na38-jp} 
at SPS,
in the Drell-Yan process and charmonium production at
Fermilab \cite{e772-dy,e866-jp}
and later on at larger RHIC energies by
measurements of high-$p_T$ particles
in $d+Au$ collisions
\cite{brahms,star}.

Althought 
forward rapidity region at RHIC
allows to investigate small-$x$
coherence phenomena, one should be
carreful with interpretation
of observed suppression. Such a suppression
is arisen globally for any reaction studied so far
at any energy. 
Namely, all fixed target experiments
have too low
energy for the onset of coherence effects since  
$x_2$ is not small. The rise of suppression with $y$   
(with Feynman $x_F$) shows the same pattern as observed
at RHIC.
 
This universality of suppression favors also
another mechanism which should be common for all reactions studied
at any energy. Such a mechanism 
was proposed in \cite{knpsj-05} and allows to describe
a strong suppression via
energy conservation effects in initial state parton rescatterings.
It can be also interpreted alternatively as a parton effective energy loss
proportional to initial energy 
leading so to $x_F$ scaling of nuclear effects.

The projectile hadron can be decomposed over different
Fock states. A nucleus has a higher resolution than a proton
due to multiple interactions and so
can resolve higher Fock components containing more constituents.
Corresponding
parton distributions fall off steeper at $x\to 1$ where
any hard reaction can be treated as a large
rapidity gap (LRG) process where no particle is produced
within rapidity interval $\Delta y = -\ln(1-x)$.   
The suppression factor as a survival probability for LRG
was estimated in \cite{knpsj-05}, $S(x)\sim 1-x$.
Each of multiple interactions of projectile partons
produces an extra $S(x)$ and the weight factors are
given by the AGK cutting rules \cite{agk}.
%
Then the effective 
parton distribution correlates with the nuclear target
\cite{knpsj-05,prepar1},
%
%
\vspace*{-0.2cm}
 \BA
f^{(A)}_{q/N}\bigl (x, 
Q^2,{\vec b}\bigr
) =
C\,f_{q/N}\bigl (x,
Q^2\bigr
)\,
exp\biggl[ -[1 - S(x)]\,\sigma_{eff}T_A({\vec b})\biggr]\, ,
\label{10}
 \EA
%
%
where $T_A({\vec b})$ is the nuclear thickness function defined
at nuclear impact parameter ${\vec b}$,
$\sigma_{eff} = 20\,$mb \cite{knpsj-05} and
the normalization
factor $C$ is fixed by the Gottfried sum rule.
  
In this paper we study a suppression of
the Drell-Yan (DY) process on a nucleus with
respect to a nucleon target and 
the rise of this suppression with $y$ ($x_1$, $x_F$) 
in various kinematic regions. First we compare our
predictions 
with data from the fixed target E772 experiment at
FNAL \cite{e772-dy}. Then similar nuclear effects are
predicted also for
the RHIC forward region expecting the same suppression pattern
as seen at FNAL. Finally we perform for the first time
predictions
in the kinematic range corresponding to a new E906
experiment planned at FNAL 
where no coherence effects are expected.
\vspace*{-0.2cm}

%
\section{The color dipole approach\label{dipole}}
%

The DY process in the target rest frame can
be treated as radiation of a heavy photon/dilepton
by a projectile quark.
The transverse momentum $p_T$ distribution of photon
bremsstrahlung in quark-nucleon interactions,
$\sigma^{qN}(\alpha,\vec{p}_T)$,
reads \cite{kst1}:
%
\vspace*{-0.2cm}
 \BA
&&
\hspace*{-.8cm}
\frac{d\sigma(qN\rightarrow \gamma^*\,X)}{d(ln\,\alpha)\,d^2p_T}
=
\frac{1}{(2\pi)^2}\,
\sum\limits_{in,f}\,
\int\,d^2 r_1\,d^2 r_2\,
e^{i\vec{p}_T\cdot(\vec{r}_1 - \vec{r}_2)}  
\Phi_{\gamma^* q}^*(\alpha,\vec{r}_1)
\Phi_{\gamma^* q}(\alpha,\vec{r}_2)\,
\Sigma(\alpha,r_1,r_2)
\label{20}
\vspace*{-0.2cm}
 \EA
%
where 
$\Sigma(\alpha,r_1,r_2) = \bigl \{\sigma_{\bar qq}(\alpha r_1)
+ \sigma_{\bar qq}(\alpha r_2)
- \sigma_{\bar qq}(\alpha |\vec{r}_1 - \vec{r}_2|)
\bigr \}/2$,
$\alpha = p^+_{\gamma^*}/p^+_q$ and
the light-cone (LC) wave functions of the projectile
$q+\gamma^*$ fluctuation $\Phi_{\gamma^* q}^*(\alpha,\vec{r})$ are
presented in \cite{kst1}.
Feynman variable is given as $x_F = x_1 - x_2$ and
in the target rest frame $x_1 = p^+_{\gamma^*}/p^+_p$.
For the dipole cross section $\sigma_{\bar qq}(\alpha r)$ in Eq.~(\ref{20})
we used GBW \cite{kmw-06} and KST \cite{kst2}
parametrizations.

The hadron cross section is given convolving the
parton cross section, Eq.~(\ref{20}), with
the corresponding parton distribution functions (PDFs)
$f_{q}$ and $f_{\bar{q}}$
\cite{kst1,krt-01},
%
\vspace*{-0.2cm}
 \BA
\frac{d\sigma(pp\rightarrow \gamma^* X)}{dx_F\,d^2p_T\,dM^2}
=
\frac{\alpha_{em}}{3\,\pi\,M^2}
\frac{x_1}{x_1 + x_2}
\int_{x_1}^{1}
\frac{d\alpha}{\alpha^2} 
\sum_q Z_q^2
\biggl\{\hspace*{-0.07cm}
f_{q}\bigl (\frac{x_1}{\alpha},Q^2\bigr )
+ f_{\bar{q}}\bigl (\frac{x_1}{\alpha},Q^2\bigr )
\hspace*{-0.10cm}
\biggr\}
\frac{d\sigma(qN\to\gamma^*X)}{d(ln\,\alpha)\,d^2p_T}
\hspace*{-0.00cm} ,
\label{30}
 \EA
%
where
$Z_q$ is the fractional quark charge,
PDFs $f_q$ and $f_{\bar q}$ are used
with the lowest order (LO) parametrization
from \cite{grv98} at the scale
$Q^2 = p_T^2 + (1 - x_1) M^2$
and the factor $\alpha_{em}/(3\pi\,M^2)$
accounts for decay of the photon into a dilepton.
\vspace*{-0.2cm}

\section{Dilepton production on 
nuclear targets}
\label{dpA} 

The rest frame of the nucleus is very convenient for study
of coherence effects.
The dynamics of the DY process is controlled by the
coherence length,
%
\vspace*{-0.2cm}
\BE
l_c
= \frac{2E_q\,\alpha(1-\alpha)}
{(1 - \alpha)\,M^2 + \alpha^2\,m_q^2 + p_T^2}
=  
\frac{1}{m_N\,x_2}
\frac{(1 - \alpha)\,M^2}{(1 - \alpha)\,M^2 + m_q^2\,\alpha^2 + p_T^2}
\ ,
\label{40}
\EE
%
where $E_q = x_q s/2m_N$ and $m_q$ is the energy and mass  
of the projectile quark 
and the center of mass energy squared
$s = (M^2 + p_T^2)/x_1 x_2$.
The fraction of the proton momentum
$x_q$ carried by the quark is related to
$x_1$ as $\alpha x_q = x_1$.\\
%
%
%
%
The coherence length is
related to the longitudinal momentum transfer, $q_L = 1/l_c$,
which controls the interference between amplitudes of the
hard reaction occurring on different nucleons.
The condition for the onset of shadowing in a hard
reaction is sufficiently {\bf long coherence
length} (LCL) in comparison with the nuclear radius, $l_c\gsim R_A$,
Here the special advantage of the color
dipole approach allows to incorporate
nuclear shadowing effects via a simple eikonalization
of $\sigma_{\bar qq}(x,r)$
\cite{zkl},
i.e. replacing $\sigma_{\bar qq}(x,r)$
in Eq.~(\ref{20}) by $\sigma_{\bar qq}^A(x,r)$:
%
\vspace*{-0.2cm}
\BE
\sigma_{\bar qq}^A =
2 \int d^2 b\,\biggl\{1 - \biggl [1 -
\frac{1}{2\,A}\,\sigma_{\bar qq}\,T_A(b)
\biggr ]^A\biggr\}\, .
\label{50}
\EE
%
The corresponding predictions for nuclear broadening
in DY reaction based on the theory \cite{kst1}
for LCL limit were presented in \cite{krtj-03}. \\
%
%
%
%
In the {\bf short coherence length} (SCL) regime
the coherence length is shorter than the mean internucleon
spacing, $l_c\lsim 1\div 2\,$fm.
In this limit there is no shadowing due to very short
duration of the $\gamma^*+q$ fluctuation.
The corresponding theory for description of the
quark transverse momentum broadening
can be found in \cite{jkt,jks-07}.

In this regime
the transverse momentum distribution for an incident proton
can be obtained integrating over $\alpha$ similarly as in Eq.~(\ref{30}):
%
\vspace*{-0.2cm}
 \BA
\frac{d\sigma(pA\rightarrow \gamma^*\,X)}{d\,x_F\,d^2p_T\,dM^2}
=
\frac{\alpha_{em}}{3\,\pi\,M^2}
\frac{x_1}{x_1 + x_2}
\int_{x_1}^{1}
\frac{d\alpha}{\alpha^2}
\sum_q Z_q^2
\biggl\{f_{q}\bigl (\frac{x_1}{\alpha},Q^2 \bigr )
+ f_{\bar{q}}\bigl (\frac{x_1}{\alpha},Q^2 \bigr )\biggr\}\,
\sigma^{qA}(\alpha,p_T)\, ,
\label{250}
 \EA
%
where
$\sigma^{qA}(\alpha,p_T)$ represents the cross section 
for an incident quark to 
produce a photon on a nucleus $A$ with transverse momentum
$p_T$. This cross section
can be expressed convolving the probability function
$W^{qA}(\vec{k}_T,x_q)$ with the cross section
$\sigma^{qN}(\alpha,k_T)$ (see Eq.~(\ref{20})),
%
\vspace*{-0.2cm}
\BE
\sigma^{qA}(\alpha,p_T)
=
\int d^2 k_T W^{qA}(\vec{k}_T,x_q) \sigma^{qN}
(\alpha,\vec{l}_T)\, ,
\label{240}
\EE
%
where $\vec{l}_T = \vec{p}_T - \alpha \vec{k}_T$.

Probability distribution in Eq.~(\ref{240}) that a quark will
acquire transverse momentum $\vec{k}_T$
on the nucleus, $W^{qA}(\vec{k}_T,x_q)$,
is obtained by the averaging procedure
over the nuclear  
density $\rho_A(b,z)$:
%
\vspace*{-0.2cm}
\BA
W^{qA}(\vec{k}_T,x_q) =
\frac{1}{A} \int d^2 b dz \rho_A(b,z)\,
W_A^q(\vec{k}_T,x_q,\vec{b},z)\, ,
\label{210}
\EA
%
where $W_A^q(\vec{k}_T,x_q,\vec{b},z)
= dn_q/d^2 k_T$ means now
the partial probability distribution that a valence quark arriving
at the position $(\vec{b},z)$ in the nucleus $A$ will have
acquired transverse momentum $\vec{k}_T$.
It can be written in term of the quark density matrix,
$\Omega_{q}(\vec{r}_1,\vec{r}_2) = (b_0^2/\pi)\,
\exp( - b_0^2 (r_1^2 + r_2^2)/2)$,
%
\vspace*{-0.2cm}
\BA
W_A^q(\vec{k}_T,x_q,\vec{b},z)
=
\hspace*{-0.1cm}
\frac{1}{(2 \pi)^2}
\int d^2 r_1 d^2 r_2\,e^{i\,\vec{k}_T\cdot
(\vec{r}_1 - \vec{r}_2)}
\Omega_{q}(\vec{r}_1,\vec{r}_2)\,
e^{- \frac{1}{2}\,\sigma_{\bar qq}
(x_q,\vec{r}_1 - \vec{r}_2)\,
T_A(\frac{\vec{r}_1 + \vec{r}_2}{2} + \vec{b},z)}
\, ,
\label{180}
\EA
%
where $b_0^2 =
\frac{2}{3\,\la r_{ch}^2\ra}$ 
with $\la r_{ch}^2\ra = 0.79\pm 0.03\,$fm$^2$ representing the
mean-square charge radius of the proton.
$T_A(b,z)$ in Eq.~(\ref{180}) is
the partial nuclear thickness function,
$T_A(b,z) = \int_{-\infty}^{z}\,dz'\,\rho_A(b,z')$.

Nuclear effects in $p+A$ collisions are usually investigated
via the so called nuclear
modification factor, defined as
$R_A(p_T,x_F,M) =
\frac{d\sigma(pA\rightarrow \gamma^*\,X)}{d\,x_F\,d^2p_T\,dM^2}/
A\,\,
\frac{d\sigma(pN\rightarrow \gamma^*\,X)}{d\,x_F\,d^2p_T\,dM^2}$,
%
where the numerator
is calculated in SCL and LCL regimes as described above.
Corrections for the finite coherence length was realized
by linear interpolation using nuclear longitudinal formfactor
\cite{knst-02} (for more sophisticate
Green function method see \cite{kst1,n-03}).

Note that at RHIC energy and at forward rapidities 
the eikonal formula for LCL regime, Eqs.~(\ref{30}) and (\ref{50}),
is not exact since higher Fock components containing gluons
lead to additional corrections, called gluon shadowing (GS).
The corresponding suppression factor $R_G$ was derived
in \cite{knst-02,krtj-03} and included in calculations
replacing in Eq.~(\ref{50}) $\sigma_{\bar qq}$
by $R_G\,\sigma_{\bar qq}$. GS leads to reduction of the
Cronin effect \cite{knst-01} at medium-high $p_T$ and to
additional suppression (see Fig.~\ref{rhic}).

In the fixed target FNAL energy range,
for elimination of the coherence
effects one can study production of dileptons at large
$M$ (see Eq.~(\ref{40})) as has been realized
by the E772 Collaboration \cite{e772-dy}.
Another possibility is to study the DY process at large 
$x_1\to 1$, when also $\alpha\to 1$, and $l_c\to 0$
in this limit (see Eq.~(\ref{40})).
\begin{figure}[htb]
\includegraphics[width=21.8pc]{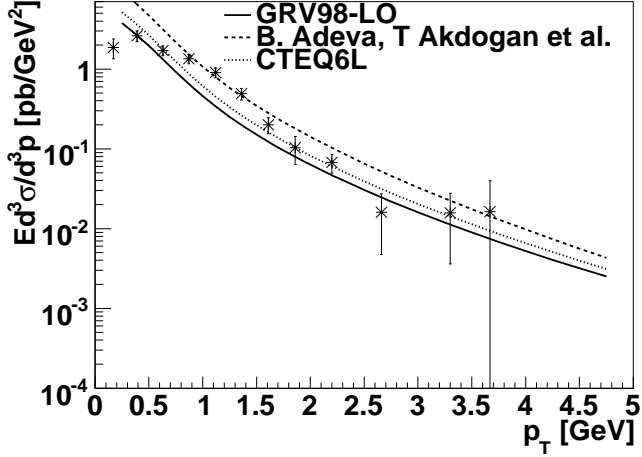}\hspace{0.0pc}%
 \begin{minipage}[b]{16pc}\caption{\label{e866-pp}
Differential cross section of dileptons in $p+p$ collisions
at $x_F = 0.63$ and $M = 4.8\,$GeV vs. E866 data \cite{e866-pp}.
}
\end{minipage}
\end{figure}
%
%
%

%
\section{Nuclear suppression at forward rapidities: model vs. data}
%

We start with
the DY process in $p+p$ collisions.
Besides calculations based on Eq.~(\ref{30}) using
GRV98 PDFs \cite{grv98} 
(see the solid line
in Fig.~\ref{e866-pp}) we present by the dashed and dotted 
line
also predictions using proton structure functions
from \cite{smc} and CTEQ6L parametrization of PDFs from 
\cite{cteq6l}, respectively.
Fig.~\ref{e866-pp} shows
a reasonable agreement of the model
with data from the E866/NuSea Collaboration \cite{e866-pp}.
This encourages us to apply the color dipole
approach to nuclear targets as well.
 \begin{figure}[htb]
\vspace*{-0.50cm}
\includegraphics[scale=0.82]{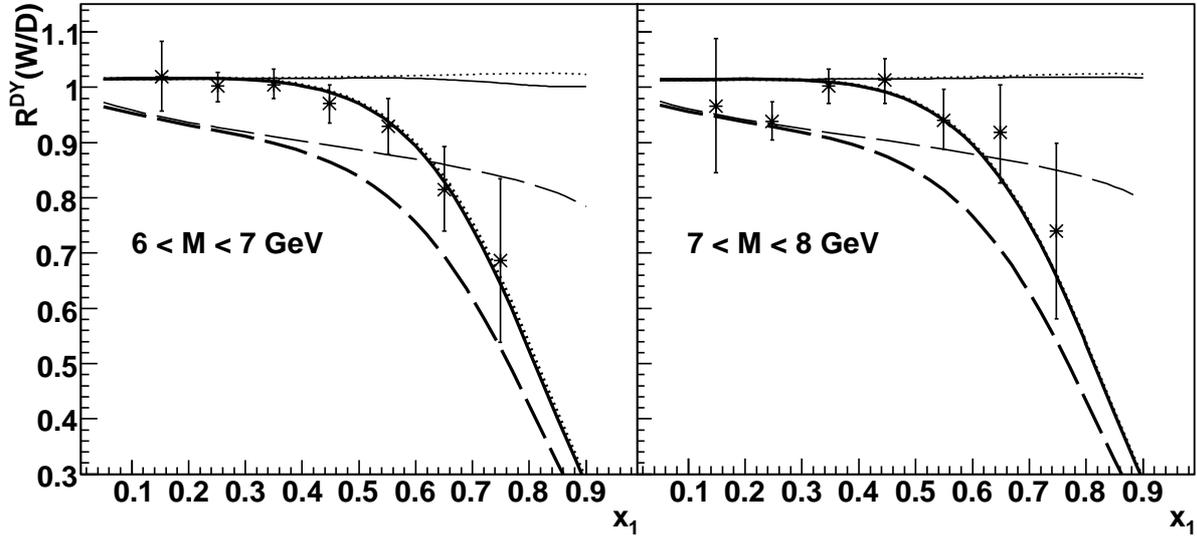}
\begin{center}

\vspace*{-0.5cm}
\caption
{
Ratio $R^{DY}(W/D)$ of Drell-Yan cross sections
on W and D vs. E772 data
for $6 < M < 7\,$GeV (Left) and $7 < M < 8\,$GeV
(Right).
Predictions correspond to SCL (dotted curves), LCL
(dashed curves) regimes and 
their interpolation (solid curves).
Thick and thin curves
are calculated with and without 
corrections (\ref{10}) for energy conservation,
respectively.
}
\label{e772}
\vspace*{-0.8cm}
\end{center}

\end{figure}


The E772 Collaboration \cite{e772-dy} found a significant suppression of
DY pairs at large $x_1$ (see Fig.~\ref{e772}).
Large invariant masses of the photon allows to minimize
shadowing effects (see a small differences between
dotted and solid lines in Fig.~\ref{e772}).
If effects of energy
conservation, Eq.~(\ref{10}), are not included one can not describe a strong
suppression at large $x_1$.
In the opposite case a reasonable agreement 
of our model with data is achieved.

\begin{figure}[htb]
\vspace*{-0.40cm}
\includegraphics[scale=0.82]{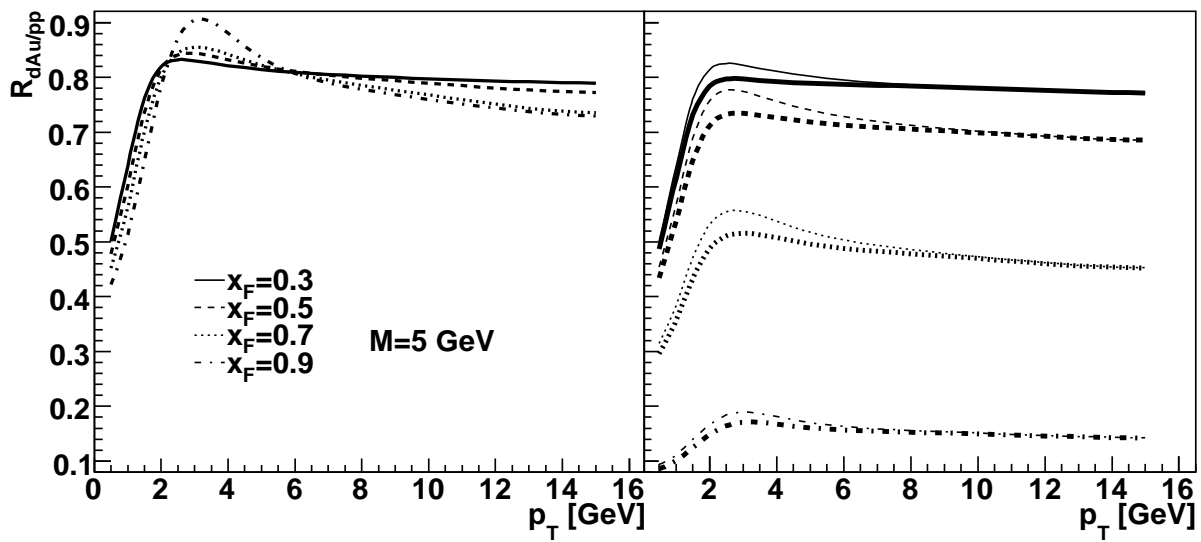}
\begin{center}

\vspace*{-0.6cm}
\caption
{(Left)
Predictions for the ratio $R_{d+Au}(p_T)$ at $\sqrt{s} = 200\,$GeV
for several fixed values of $x_F$ without effects of effective energy
loss.
(Right) The same as (Left) but
with effects of effective energy
loss, Eq.~(\ref{10}), (thin lines).
Thick lines
additionally include GS effects.
}
\label{rhic}
\end{center}
\vspace*{-0.8cm}

\end{figure}

One can approach the kinematic limit increasing
$p_T$. 
Therefore
we present also predictions for $p_T$ dependence
of the nuclear modification factor $R_{d+Au}$ at RHIC energy
and at several fixed values of $x_F$.
Similarly as in \cite{knpsj-05} instead of usual Cronin enhancement, a
suppression is found (see Fig. ~\ref{rhic}).
The onset of isotopic effects (IE) in $d+Au$ collisions at large $p_T$ gives the values
$R_{d+Au}^{IE}\sim 0.73\div 0.79$ depending on $x_F$.
In $p + Au$ collisions the corresponding ratio $R_{p+Au}\to 1$ from above and
no nuclear effects are assumed at large $p_T$ expecting so QCD
factorization.
However, we predict a strong onset of effective energy loss effects, Eq.~(\ref{10}), at large $x_F$
(see Fig.~\ref{rhic})
quantifying itself as a large deviation of suppression from the above values
$R_{d+Au}^{IE}$.
The predicted huge rise of suppression
with $x_F$ in Fig.~\ref{rhic} reflects much smaller survival probability
$S(x_F)$ at larger $x_F$ and can be tested in the future
by the new data from RHIC.
Note that effects of GS depicted in Fig.~\ref{rhic}
by the thick lines lead to additional suppression
which rises with $x_F$.

 \begin{figure}[htb]
\vspace*{-0.70cm}
\includegraphics[scale=0.81]{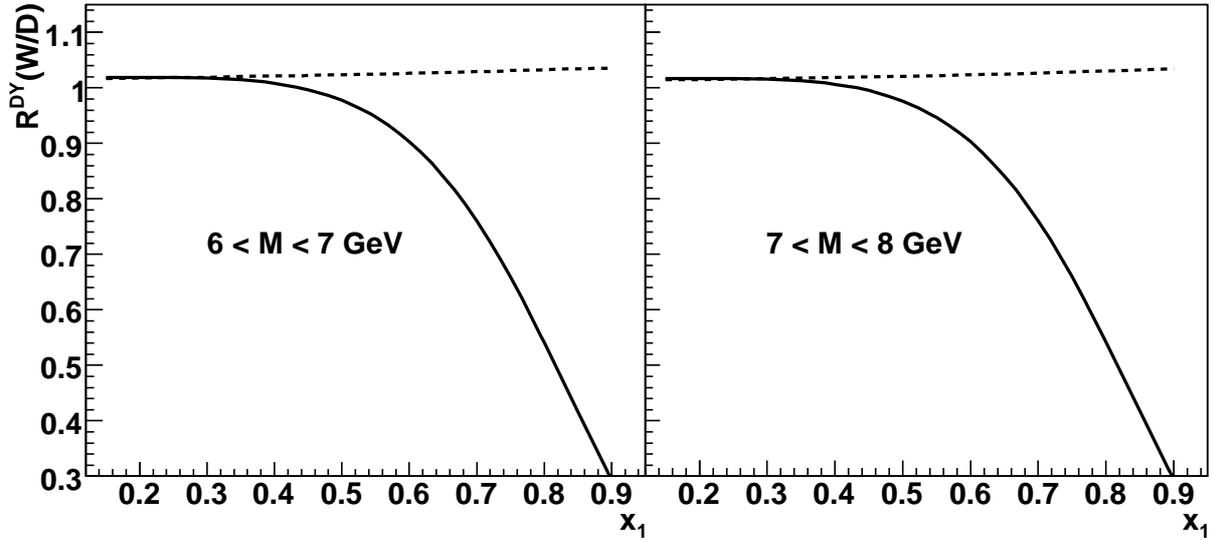}
\begin{center}

\vspace*{-0.5cm}
\caption
{(Left)
Predictions for the
ratio $R^{DY}(W/D)$ of Drell-Yan cross sections
on W and D
for $6 < M < 7\,$GeV (Left) and $7 < M < 8\,$GeV
(Right)
realized for the kinematic range of the planned
E906 experiment at Fermilab.
Solid and dashed curves are calculated with and without
effects of effective energy loss, Eq.~(\ref{10}), respectively.
}
\label{e906}
\end{center}
\vspace*{-1.2cm}

\end{figure}


Finally we present in Fig.~\ref{e906} for the first time predictions
for $x_1$ dependence of the nucleus-to-nucleon ratio in the kinematic
range corresponding to a new E906 experiment planned
at Fermilab. We shoud not expect any shadowing effects since
initial energy is small, $E_{lab} = 120\,$GeV and a strong
nuclear suppression at large $x_1$ is caused predominantly
by the energy conservation constraints, Eq.~(\ref{10}).
\vspace*{-0.4cm}

%
%
\section{Summary}
%
%

We demonstrate that  
besides an onset of coherence 
a nuclear suppression at forward rapidities 
(large $x_1$, $x_F$) 
can be induced also by energy conservation effects in multiple parton
rescatterings
interpreted alternatively as a parton effective energy loss
proportional to initial energy.
Universality of this treatment is in its applicability to
any reaction studied at any energy also in the kinematic
regions where coherence phenomena (shadowing, CGC) 
can not be manifested.
First we apply this approach to the DY process
and explain well a significant suppression at large $x_1$
in accordance with the E772 data. The FNAL energy range and
large invariant masses of the photon allow to minimize
the effects of coherence, what does not leave much room
for other mechanisms, such as CGC.
Then 
we predict a significant suppression also for $d+Au$ collisions
at RHIC in the forward region (see Fig.~\ref{rhic}).
At small 
$p_T$ we show an
importance of GS effects and their rise with $x_F$.
Finally
we 
present for the first time predictions for
strong nuclear effects expected in a new E906
experiment planned at FNAL. Much smaller beam
energy than in E772 experiment allows to exclude safely
interpretations based on coherence phenomena.
\vspace*{-0.3cm}

\ack
This work was supported in part by the
Slovak Funding Agency, Grant 2/0092/10 and by Grants VZ M\v SMT
6840770039 and LC 07048 (Ministry of Education of the Czech Republic).

\end{document}